\documentclass[aps,twocolumn,showpacs,preprintnumbers,amsmath,amssymb,superscriptaddress,floatfix,nofootinbib]{revtex4}

\usepackage{graphicx}
\usepackage{epsfig}
\usepackage{epstopdf}
\usepackage{subfigure}

\usepackage{amsmath}
\usepackage{amsfonts}
\usepackage{amssymb}
\usepackage{color}
\usepackage[colorlinks,
            citecolor=blue,
            anchorcolor=red,
            menucolor=red,
            linkcolor=red,
            filecolor=red,
            runcolor=red,
            urlcolor=blue,
            frenchlinks=red]{hyperref}

\begin{document}
\title{The mass spectrum and strong decay properties of the charmed-strange mesons within Godfrey-Isgur model considering the coupled-channel effects}

\author{Jing-Jing Yang}
\affiliation{School of Physics and Microelectronics, Zhengzhou University, Zhengzhou, Henan 450001, China}
  

\author{Xiaoyu Wang}
\affiliation{School of Physics and Microelectronics, Zhengzhou University, Zhengzhou, Henan 450001, China}\vspace{0.5cm}
	
\author{De-Min Li}
\affiliation{School of Physics and Microelectronics, Zhengzhou University, Zhengzhou, Henan 450001, China}\vspace{0.5cm}
 
\author{Yu-Xiao Li}
\affiliation{School of Physics and Microelectronics, Zhengzhou University, Zhengzhou, Henan 450001, China}\vspace{0.5cm}

\author{En Wang}
\email{wangen@zzu.edu.cn}
\affiliation{School of Physics and Microelectronics, Zhengzhou University, Zhengzhou, Henan 450001, China}\vspace{0.5cm}
\affiliation{Guangxi Key Laboratory of Nuclear Physics and Nuclear Technology, Guangxi Normal University, Guilin 541004, China}
\date{\today}

\author{Wei Hao}
\email{haowei@nankai.edu.cn}
\affiliation{School of Physics, Nankai University, Tianjin 300071, China}

\begin{abstract}
Motivated by the recently observed $D_{s0}(2590)$ state by LHCb, we investigate the mass spectrum and the strong decay properties of the charmed-strange mesons within the Godfrey-Isgur model considering the coupled-channel effects. One finds that the  $D^*K^*$ contributions to the mass shifts are large for all the charmed-strange states, which is maybe due to the spin-enhancement effect.
Our results support that $D_{s0}^*(2317)$ and $D_{s1}(2460)$ could be interpreted as the $D_{s}(1^3P_0)$ and $D_{s}(1^3P_1)$ states with larger $DK$ and $D^*K$ components, respectively, although the probabilities of the $DK$ and $D^*K$ components for $D_{s0}^*(2317)$ and $D_{s1}(2460)$ are smaller than other theoretical predictions, which may be due to our neglect of the direct interaction of the meson components.
Meanwhile, $D_{s1}(2700)$, $D_{s1}(2536)$, $D^*_{s2}(2573)$, $D_{s1}^*(2860)$, $D_{s3}^*(2860)$, and $D_{sJ}^*(3040)$ could be well interpreted as the $D_s(2^3S_1)$, $D_s(1^1P_1)$, $D_s(1^3P_2)$, $D_s(1^3D_1)$, $D_s(1^3D_3)$, and $D_s(2^1P_1)$ states, respectively. Although the $D_{s0}(2590)$ mass is about 50~MeV less than our prediction for the $D_{s0}(2S)$ state, its width is still in good agreement with the one of $D_{s0}(2S)$. Therefore, $D_{s0}(2590)$ state needs to be further confirmed by the experimental measurements, and the more precise information about $D_{s0}(2590)$ will shed light on its assignment of $D_{s0}(2S)$. Furthermore, we predict the masses and the strong decay properties of the charmed-strange mesons with masses around 3~GeV, which would be helpful to experimentally search for these states.      
\end{abstract}

\maketitle

\section{Introduction}
Recently, LHCb Collaboration observed a new excited $D_{s0}(2590)$ state with mass $M=2591\pm6\pm7$~MeV and width $\Gamma=89\pm16\pm12$~MeV in the $D^{+}K^{+}\pi^{-}$ mass distribution of the $ B^{0} \rightarrow D^{-}D^{+}K^{+}\pi^{-}$ decay using a data sample with integrated luminosity of 5.4 $ fb^{-1} $ at centre-of-mass energy of 13~TeV, and its spin-parity was determined to be $J^P=0^-$~\cite{LHCb:2020gnv}. According to the Review of Particle Physics (RPP)~\cite{ParticleDataGroup:2022pth}, there are several charmed-strange mesons, which contain the $D_s$, $D^*_s$, $D_{s0}^*(2317)$, $D_{s1}(2536)$, $D_{s1}(2460)$, $D_{s2}^*(2573)$, $D_{s1}^*(2860)$, $D_{s3}^*(2860)$, $D_{s1}^*(2700)$, and $D_{sJ}(3040)$, and there have been many studies about the charmed-strange mesons~\cite{Rosner_2007,Segovia:2015dia,Chen:2016spr,Song:2015nia,Song:2014mha,Li:2009qu,Li:2007px}. Although the newly observed $D_{s0}(2590)$ was suggested to be the candidate of the $D_{s0}(2S)$ state by LHCb~\cite{LHCb:2020gnv}, it still draws particular attention on the spectrum of the charmed-strange mesons due to the fact that the mass of the observed $D_{s0}(2590)$ is about 80~MeV less than the $D_{s0}(2S)$ mass predicted by the conventional quark models~\cite{Ni:2021pce,Ebert:2009ua,Zeng:1994vj,Godfrey:2015dva,Godfrey:1985xj}. 

In Ref.~\cite{Wang:2021orp}, the authors have investigated the mass and the strong decay width of $D_{s0}(2590)$, and concluded that $D_{s0}(2590)$ was hardly  interpreted as the $D_{s0}(2S)$ state. In Ref.~\cite{Xie:2021dwe}, it is shown that the $P$-wave $D^*K$ interaction plays an important role to obtain the mass and width of $D_{s0}(2590)$. In Ref.~\cite{Ortega:2021fem}, $D_{s0}(2590)$ can be regarded as a $D_{s0}(2S)$ state plus the important effect of the nearby meson-meson thresholds by performing a coupled-channel calculation including the $ D^{(*)}K^{(*)}$, $D^{(*)}_s \omega $, and $ D^{(*)}_s\eta $ channels. In Ref.~\cite{Gao:2022bsb},  the $D_{s0}(2590)$ is studied within the Godfrey-Isgur (GI) relativistic quark model including screening effects and the $^{3}P_{0} $ model, which supports the interpretation as the $D_{s0}(2S)$. In addition, Ref.~\cite{Hao:2022vwt} has made a systematic calculation of the spectrum and strong decays of the charmed-strange system in a coupled-channel framework, and the mass and width of the $D_{s0}(2590)$ could be reasonably described. 

As we known, BaBar and CLEO have observed $D_{s0}^*(2317)$ and $D_{s1}(2460)$ in the $D^+_s\pi^0$ channel~\cite{BaBar:2003oey,CLEO:2003ggt}, and their small widths of $\Gamma_{D_{s0}(2317)}<3.8$~MeV and $\Gamma_{D_{s1}(2460)}<3.5$~MeV  imply a minimal violation of the isospin conservation. In addition, the masses of the observed $D_{s0}^*(2317)$ and $D_{s1}(2460)$ resonances are much lower than the corresponding predictions from the conventional quark models \cite{Godfrey:1985xj,DiPierro:2001dwf,Godfrey:2015dva} and the Lattice QCD calculations~\cite{Bali:2003jv,Dougall:2003hv}, 
which motivates many interpretations for their structure, such as compact $[cq][\bar{s}\bar{q}]$ tetraquark, molecular states, and the mixing of the $c\bar{s}$ and other components~\cite{Guo:2017jvc,Chen:2022asf,Oset:2016lyh,Albaladejo:2018mhb,Kolomeitsev:2003ac}. In Refs.~\cite{Hofmann:2003je,Barnes:2003dj,Chen:2004dy}, $D_{s0}^*(2317)$ and $D_{s1}(2460)$ were explained as the $DK$ and $D^*K$ molecular states, respectively, which was supported by the studies of the heavy chiral unitary approach \cite{Guo:2006fu,Guo:2006rp} and the unitarized coupled channel framework~\cite{Gamermann:2006nm}. 
In Ref.~\cite{Gamermann:2007fi}, the axial resonance $D_{s1}(2460)$ could be also dynamically generated from the interactions of pseudoscalar-vector within the SU(4) flavor symmetry. 
In Ref.~\cite{Lutz:2007sk}, $D_{s0}^*(2317)$ and $D_{s1}(2460)$ could be dynamically generated by the coupled-channel dynamics based on the leading order chiral Lagrangian. In addition, the $D^*_{s2}(2573)$ was predicted to couple strongly to the $D^*K^*$ $(D^*_s \phi(\omega))$ channels under the vector-vector interaction within the hidden gauge formalism in a coupled channel unitary approach~\cite{Molina:2010tx}. Thus,  it implies that more components are necessary to describe the properties of the charmed-strange mesons.



Although the general potential models, such as GI relativistic quark model~\cite{Godfrey:1985xj}, could provide a good description for most of the meson spectra, the coupled-channel effects (or the pair-creation effects), which were usually neglected, will manifest as a coupling to meson-meson (meson-baryon) channels and lead to mass shifts. It has been shown that the coupled-channel effects play an important role for describing the mesons spectra, such as charmonium~\cite{Kalashnikova:2005ui,Li:2009ad,Ferretti:2013faa}, bottomonium~\cite{Liu:2011yp,Ferretti:2012zz,Ferretti:2013vua,Lu:2016mbb}, and charmed-strange mesons~\cite{vanBeveren:2003jv,vanBeveren:2003kd,Coito:2011qn,Hwang:2004cd,Simonov:2004ar,Lee:2004gt,Guo:2007up,Zhou:2011sp,Badalian:2007yr,Dai:2006uz}. Therefore, in this work we will investigate the mass spectrum of the charmed-strange mesons within the GI  quark model by taking into account the mass shifts from the coupled-channel effects, where the potential model parameters will be refitted. 

The rest of the paper is organized as follows. In Sec.~\ref{sec:model}, we will present our theoretical models, including the coupled-channel model and the GI relativistic quark model. In Sec.~\ref{sec:results}, the numerical results will be presented. We will conclude the work and give the summary in Sec.~\ref{sec:summary}.

\section{The theoretical models}
\label{sec:model}

\subsection{The coupled-channel model}
\label{sec:couple}

In the coupled-channel model, the Hamiltonian of a meson system is defined as \cite{Ferretti:2012zz,Ferretti:2013faa,Ferretti:2013vua,Ferretti:2015rsa}
\begin{equation}
		H=H_{0}+H_{BC}+H_{I} \label{eq:hamilton}
\end{equation}
where $H_0$ connects with the bare mass $M_0$ of the meson $A$, and is obtained from  the GI model. $H_{BC}$ is the Hamiltonian of the intermediate  mesons $B$ and $C$, which couple to the meson $A$, and $H_I$ describes the interaction of the meson state
$|A\rangle$ and the intermediate meson-meson continuum $|BC\rangle$, and  connects with the mass shifts $\Delta M$ from the coupled-channel effects.

The Hamiltonian $H_{0}$ of the GI model will give rise to the bare mass $M_0$ of the meson $A$
	\begin{eqnarray}
		H_{0}|A\rangle=M_{0}|A\rangle,
	\end{eqnarray}
and we will discuss the Hamiltonian $H_{0}$ of the GI model in next subsection. As done in Refs.~\cite{Hao:2022vwt,Xie:2021dwe,Kalashnikova:2005ui,Lu:2016mbb,Ferretti:2013vua}, we assume that there is no interaction between the $BC$ pair, and only the kinetic energy of the intermediate $BC$ pair will be considered\footnote{Indeed, it is a systematic way to include  the interactions of the meson-meson consistently at the quark level, which will  introduce some free parameters and should be complicated, since we try to describe the mass spectrum of the charmed-strange mesons considering the coupled-channel effects of pseudoscalar-pseudoscalar, pseudoscalar-vector, and vector-vector. One will find that the present model could give a reasonable description for the mass spectrum of the charmed-strange mesons, thus we neglect the interactions of the meson-meson, as done in Refs.~\cite{Hao:2022vwt,Xie:2021dwe,Kalashnikova:2005ui,Lu:2016mbb,Ferretti:2013vua}.}. The Hamiltonian $H_{BC}$ can be written as the sum of the kinetic energies of $B$ and $C$, and the Schr\"odinger equation for the $BC$ pair is derived as follows
	\begin{eqnarray}
		H_{BC}|BC\rangle=E_{BC}(p)|BC; p \rangle,
	\end{eqnarray}
	\begin{eqnarray}
		E_{BC}(\boldsymbol{p})=\sqrt{m_{B}^{2}+p^{2}}+\sqrt{m_{C}^{2}+p^{2}},
	\end{eqnarray}
where $p$ is the center of mass momentum of the meson pair $BC$ running from 0 to infinity, $m_B$ and $m_C$ are the masses of the meson $B$ and $C$, respectively.

Taking into account the coupled-channel effects $H_I$, the Schr\"odinger equation with the  Hamiltonian given in Eq.~(\ref{eq:hamilton}) can be written as
	\begin{eqnarray}
		H|\psi\rangle=M|\psi\rangle,
	\end{eqnarray}
where $|\psi\rangle$ is the eigen wave function of the system, which can be expressed as
	\begin{eqnarray}
		|\psi\rangle =a_{0}|A\rangle+\sum_{BC}\int d^{3}p \, c_{BC}(p)|BC, p\rangle,
	\end{eqnarray}
where the coefficients $ a_{0} $ and $c_{BC} $ are the normalizing constants of the corresponding wave functions of $q\Bar{q}$ bare state and $BC$ component, respectively.

Thus, the physical mass $M$ in the coupled-channel model is given by
\begin{gather}
M = M_0 + \Delta M, \label{m} \\
\Delta M = \sum_{BC\ell J} \int_0^{\infty} p^2 dp \mbox{ } \frac{\left|\left\langle BC;p \right| T^\dag \left| A \right\rangle \right|^2}{M - E_{BC}+i\epsilon}, \label{deltam}
\end{gather}
where $\left\langle BC;p \right| T^\dag \left| A \right\rangle$ is the transition amplitude for the operator $T^\dag$ between the intermediate state $|BC\rangle$ and the meson $A$. $BC$ has various channels and the sum runs over all the channels we will consider in this work. $\ell$ is the orbital angular momentum, and the total angular momentum is $J=J_B+J_C+\ell$.
In our calculation, we adopt the quark-antiquark pair-creation operator $T^\dag$ from the $^3P_0$ model~\cite{Ferretti:2013faa,Ferretti:2012zz,Ferretti:2013vua}, which could be expressed as
\begin{equation}
	\label{eqn:Tdag}
	\begin{array}{rcl}
	T^{\dagger} &=& -3 \, \gamma_0^{eff} \, \int d \boldsymbol{p}_3 \, d \boldsymbol{p}_4 \, 
	\delta(\boldsymbol{p}_3 + \boldsymbol{p}_4) \,  
	{e}^{-r_q^2 (\boldsymbol{p}_3 - \boldsymbol{p}_4)^2/6 }\,  \\
	& &  C_{34}  F_{34}  \left[ \chi_{34} \, \times \, {\cal Y}_{1}(\boldsymbol{p}_3 - \boldsymbol{p}_4) \right]^{(0)}_0 \, 
	b_3^{\dagger}(\boldsymbol{p}_3) \, d_4^{\dagger}(\boldsymbol{p}_4) ~,   
	\end{array}
\end{equation}
where $C_{34}$, $F_{34}$, and $\chi_{34}$ are the color-singlet wave function, flavor-singlet wave function, and spin-triplet wave function for the created quark and antiquark pair $q\bar{q}$, respectively. $ b_3^{\dagger}(\boldsymbol{p}_3)$ and $d_4^{\dagger}(\boldsymbol{p}_4)$ are the creation operators for a quark and an antiquark with three-momenta $\boldsymbol{p}_3$ and $\boldsymbol{p}_4$, respectively.  $\gamma_0^{eff}=\frac{m_u}{m_i}\gamma_0$ ($m_i$ are the quark masses of $u$, $d$, or $s$) is the effective pair-creation strength, and in our calculation its value is obtained by fitting to the strong decay of $D_{s2}^*(2573)$, which is well interpreted as the $D_s(1^3P_2)$ state. In the $^3P_0$ model, the operator $T^{\dagger}$ creates a pair of constituent quarks with an  effective size, the pair-creation point has to be smeared out by a Gaussian factor, where $r_q$ was determined from meson decays to be in the range $0.25\sim 0.35$~fm~\cite{Silvestre-Brac:1991qqx,Geiger:1991ab,Geiger:1991qe,Geiger:1996re}. In our calculation, we take the value $r_q = 0.3$~fm.

Under the Simple Harmonic Oscillator (SHO) approximation, the meson wave function in the momentum space can be expressed as
\begin{eqnarray}
\psi_{nLM_L}^{\text{SHO}}(p)=R_{nL}^{\text{SHO}}(p)Y_{LM_L}(\Omega_p),
\end{eqnarray}
where the radial wave function is given by
\begin{eqnarray}
R_{nL}^{\text{SHO}}(p)=\frac{(-1)^n(-i)^L}{\beta^{3/2}}\sqrt{\frac{2n!}{\Gamma(n+L+3/2)}}\nonumber\\
\times\left(\frac{p}{\beta}\right)^Le^{-(p^2/2{\beta}^2)}L_n^{L+1/2}
\left(\frac{p^2}{\beta^2}\right),
\end{eqnarray}
here $\beta$ is the SHO wave function scale parameter, and $L_n^{L+1/2}\left({p^2}/{\beta^2}\right)$ is an associated Laguerre polynomial. The corresponding parameters are tabulated in Table~\ref{3p0parameter}. 

\begin{table}[htpb] 
\caption{Parameters of the coupled-channel model.}
\label{3p0parameter}
\begin{center}
\begin{tabular}{cc} 
\hline 
\hline 
Parameter  &  Value     \\ 
\hline 
$\gamma_0$ & $0.478$       \\  
$\beta$   & $0.4$ GeV   \\  
$r_q$      & $0.3$ fm    \\
$m_n$      & $0.33$ GeV   \\
$m_s$      & $0.55$ GeV   \\
$m_c$      & $1.50$ GeV    \\   
\hline 
\hline
\end{tabular}
\end{center}
\end{table}

If the mass of the initial meson $A$ is above the threshold of coupled-channel $BC$, the strong decays of $A\to B C$ will happen, and the strong decay width can be expressed as
\begin{equation}
	\Gamma_{A \rightarrow BC} = \Phi_{A \rightarrow BC}(p_0) \sum_{\ell, J} 
	\left| \left\langle BC,p_0, \ell J \right| T^\dag \left| A \right\rangle \right|^2 \mbox{ },
\end{equation}
 where $\Phi_{A \rightarrow BC}(p_0)$ is the standard relativistic phase space factor \cite{Ackleh:1996yt,Barnes:2005pb}
 
\begin{equation} 
	\Phi_{A \rightarrow BC} = 2 \pi p_0 \frac{E_B(p_0) E_C(p_0)}{m_A}  \mbox{ },
\end{equation}
depending on the relative momentum $p_0$ between $B$ and $C$ and on the energies of the two intermediate state mesons, 

\begin{gather}
p_0=\frac{\sqrt{\left[m_A^2-(m_B+m_C)^2\right]\left[m_A^2-(m_B-m_C)^2\right]} }{2m_A},\\
 E_B(p_0) = \sqrt{m_B^2 + p_0^2},\\   
E_C(p_0) = \sqrt{m_C^2 + p_0^2}.
\end{gather}

For the initial states below the threshold of the coupled-channels, the probabilities of each meson-meson continuum components can be calculated by
\begin{align}
	P_{BC} =& \left[1+\sum_{BC\ell J} \int_0^{\infty} p^2 dp \mbox{ } \frac{\left|\left\langle BC;p \right| T^\dag \left| A \right\rangle \right|^2}{(M - E_{BC})^2}\right]^{-1} \nonumber \\
    &\times \int_0^{\infty} p^2 dp \mbox{ } \frac{\left|\left\langle BC;p \right| T^\dag \left| A \right\rangle \right|^2}{(M - E_{BC})^2} .
\end{align}
The sum in the formula is for all the intermediate states we considered. And the probabilities  of the $c\bar{s}$ component can be calculated by $1-P_{BC}$.

For the coupled-channels, we consider ground state mesons, which include $DK$, $DK^*$, $D^*K$, $D^*K^*$, $D_s\eta$, $D_s\eta^\prime$, $D_s\phi$, $D_s^*\eta$, $D_s^*\eta^\prime$, and $D_s^*\phi$, as Refs.~\cite{Kalashnikova:2005ui,Ferretti:2013faa,Ferretti:2012zz,Ferretti:2013vua,Lu:2016mbb}. The physical masses $M$ and the mass shifts $\Delta M$ can be simultaneously determined from Eqs.~(\ref{m}) and (\ref{deltam}). The masses of the mesons used in this work are taken from RPP~\cite{ParticleDataGroup:2022pth}.

\subsection{GI Relativistic quark model}
As mentioned above, the bare mass $M_0$ in Eq.~(\ref{m}) is calculated by the potential model, the one used in the Godfrey-Isgur relativistic quark model in this work~\cite{Godfrey:1985xj}. In the GI model, the Hamiltonian of a meson system is defined as~\cite{Godfrey:1985xj}
\begin{eqnarray}
\tilde{H}&=&(p^2+m_1^2)^{1/2}+(p^2+m_{2}^2)^{1/2} +  \tilde{H}_{12}^{\text{conf}} \nonumber \\ 
&&+\tilde{H}_{12}^{\text{hyp}}+\tilde{H}_{12}^{\text{so}},
\label{ha}
\end{eqnarray}
where $\tilde{H}^\text{conf}_{12}$ is a spin-independent potential; $\tilde{H}^\text{hyp}_{12}$ is a color-hyperfine interaction which includes a tensor hyperfine potential $\tilde{H}^\text{tensor}_{12}$ and a contact hyperfine potential $\tilde{H}^\text{c}_{12}$; $\tilde{H}^\text{so}_{12}$ is a spin-orbit interaction which includes a vector spin-orbit potential $\tilde{H}^\text{so(v)}_{12}$ and a scalar spin-orbit potential $\tilde{H}^\text{so(s)}_{12}$.
In this subsection, we will first discuss the Hamiltonian terms in the non-relativistic limit, and then modify the terms to introduce the relativistic effects.
Hereafter, we will denote the terms with a tilde to be the ones considering relativistic effects, otherwise the terms without the tilde to be the ones in the non-relativistic limit.

In the non-relativistic limit, the spin-independent potential $\tilde{H}^\text{conf}_{12}$ of Eq.~(\ref{ha}) will be expressed as $H_{12}^{\text{conf}}$,
\begin{eqnarray}
\tilde{H}_{12}^{\text{conf}} &\to &  H_{12}^{\text{conf}}  =G(r)+S(r),  \\
 G(r)&=&\frac{\alpha_s(r)}{r} \boldsymbol{F_1} \cdot \boldsymbol{F_2}, ~~  S(r)= br+c,
 \label{hyp}
\end{eqnarray}
where $G(r)$ stands for the short-range one-gluon-exchange potential, and $S(r)$ corresponds to the long-range confinement. The parameters $b$ and $c$ are constants, and $\alpha_s(r)$ is the running coupling constant of QCD. $\boldsymbol{F}$ is related to the Gell-Mann matrix by $\boldsymbol{F}_1=\boldsymbol{\lambda}_1/2$ for quarks and $\boldsymbol{F}_2=-\boldsymbol{\lambda}^*_2/2$ for antiquarks, with
$\langle\boldsymbol{F}_1\cdot\boldsymbol{F}_2\rangle=-4/3$ for mesons.

The color-hyperfine interaction $\tilde{H}^\text{hyp}_{12}$ of Eq.~(\ref{ha}) could be expressed as $H^{\text{hyp}}_{12}$ in the non-relativistic limit
\begin{eqnarray}
\tilde{H}_{12}^{\text{hyp}} &\to& H^{\text{hyp}}_{12} \nonumber \\
&=&-\frac{\alpha_s(r)}{m_1m_2}\Bigg[\frac{8\pi}{3}\boldsymbol{S}_1\cdot\boldsymbol{S}_2\delta^3 (\boldsymbol r)+\frac{1}{r^3}\nonumber\\
&&\Big(\frac{3\boldsymbol{S}_1\cdot\boldsymbol{r} \boldsymbol{S}_2\cdot\boldsymbol{r}}{r^2}
 -\boldsymbol{S}_1\cdot\boldsymbol{S}_2\Big)\Bigg] \boldsymbol{F_1} \cdot \boldsymbol{F_2}.
\end{eqnarray}

The spin-orbit interaction $\tilde{H}^{\text{so}}_{12}$ of Eq (\ref{ha}) will be expressed as $H^{\text{so}}_{12}$ in the non-relativistic limit,
\begin{eqnarray}
\tilde{H}_{12}^{\text{so}} \to H^{\text{so}}_{12}=H^{\text{so(cm)}}_{12}+H^{\text{so(tp)}}_{12},
\end{eqnarray}
where $H^{\text{so(cm)}}_{12}$ is the
color-magnetic term, and $H^{\text{so(tp)}}_{12}$ is the Thomas-precession term, i.e.,
\begin{eqnarray}
H^{\text{so(cm)}}_{12}&=&-\frac{\alpha_s(r)}{r^3}\left(\frac{1}{m_1}+\frac{1}{m_2}\right)\left(\frac{\boldsymbol{S}_1}{m_1}+\frac{\boldsymbol{S}_2}{m_2}\right) \cdot \boldsymbol{L}(\boldsymbol{F}_1\cdot\boldsymbol{F}_2), \nonumber \\
\end{eqnarray}
\begin{eqnarray}
H^{\text{so(tp)}}_{12}=\frac{-1}{2r}\frac{\partial H^{\text{conf}}}{\partial
r}\Bigg(\frac{\boldsymbol{S}_1}{m^2_1}+\frac{\boldsymbol{S}_2}{m^2_2}\Bigg)\cdot \boldsymbol{L}.
\end{eqnarray}

In the above expressions, $\boldsymbol{S}_1$ and $\boldsymbol{S}_2$ denote the spin of the quark and antiquark, respectively, and $\boldsymbol{L}$ is the orbital angular momentum between the quark and antiquark.

In the GI model, the relativistic effects are introduced in two ways. Firstly, the smearing transformation is used for the non-relativistic potentials $G(r)$ and $S(r)$,
\begin{eqnarray}
    \tilde{f}(r)=\int f(r)  \rho(\boldsymbol{r}-\boldsymbol{r}')d^3\boldsymbol{r'},  \\
     \rho(\boldsymbol{r}-\boldsymbol{r}')=\frac{\sigma^3}{\pi^{3/2}}e^{-\sigma^2(\boldsymbol{r}-\boldsymbol{r}')^2}.
\end{eqnarray}
where $\rho(\boldsymbol{r}-\boldsymbol{r}')$ is a smearing function, $\sigma$ is a  parameter, and the $f(r)$ represents $G(r)$ and $S(r)$.

Secondly, since the reflection of relativistic effects lies in the momentum dependence
of interactions between quark and anti-quark, the potentials will be modified by the momentum-dependent factor as
\begin{eqnarray}
\tilde{G}(r) &\to& \left(1+\frac{p^2}{E_1 E_2}\right)^{1/2} \tilde{G}(r) \left( 1+\frac{p^2}{ E_1 E_2}\right)^{1/2}, \\
\frac{\tilde{V}_i(r)}{m_1 m_2} &\to& \left( \frac{m_1 m_2}{E_1 E_2}\right)^{1/2+\epsilon_i}  \frac{\tilde{V}_i(r)}{m_1 m_2} \left( \frac{m_1 m_2}{E_1 E_2}\right)^{1/2+\epsilon_i},\nonumber \\ \label{eqGV}
\end{eqnarray}
where $E_1=(p^2+m^2_1)^{1/2}$, $E_2=(p^2+m^2_2)^{1/2}$ are the energies of the quark and antiquark in
the mesons. The index $i$ in the parameters $\tilde{V}_i(r)$ and $\epsilon_i$ corresponds to different types of interaction in Eq. (\ref{ha}), including $i=$ contact(c), tensor(t), vector spin-orbit[so(v)] and scalar spin-orbit[so(s)] potentials.
The details of these  effective potentials can be found in Ref.~\cite{Godfrey:1985xj}.

\section{Results and discussions}
\label{sec:results}

\begin{table}[htpb] 
\caption{Fitted parameters of Godfrey-Isgur model.} 
\label{giparameter}
\begin{center}
\begin{tabular}{cc} 
\hline 
\hline 
Parameter  &  Value     \\ 
\hline 
$b$                   & $0.1614$ GeV$^2$      \\  
$c$                   & $0.0725$ GeV   \\  
$\sigma_0$            & $3.2666$ GeV    \\
$s$                   & $2.4980$   \\
$\epsilon_c$          & $-0.0788$    \\
$\epsilon_t$          & $0.6443$     \\
$\epsilon_{\mathrm{so(v)}}$      & $-0.2511$    \\
$\epsilon_{\mathrm{so(s)}}$      & $0.9001$    \\  
\hline 
\hline
\end{tabular}
\end{center}
\end{table}

\begin{table*}[htpb]
\begin{center}
\caption{\label{tab:dsm} The mass spectrum (in MeV) of charmed-strange mesons. Column 5 shows our predicted masses. Columns 6 and 7 show the GI model results and experimental values, respectively.}
\footnotesize
\begin{tabular}{ccccccc}
\hline\hline
  $n^{2S+1}L_J$  & states               &$M_0$  &$\Delta M$  &$M~(\mathrm{this~work})$     &GI \cite{Godfrey:1985xj}   & PDG~\cite{ParticleDataGroup:2022pth}  \\\hline
  $1^1S_0$     & $D_{s}$              &$2163$   &$-195$     &$1968$   &$1960$   &$1968.34\pm0.07$    \\
  $1^3S_1$     & $D_{s}^{*}$          &$2334$   &$-221$     &$2112$   &$2130$   &$2112.2\pm0.4$    \\
  $2^1S_0$     & $D_{s0}(2590)$       &$2859$   &$-213$     &$2646$   &$2670$   &$2591\pm6\pm7$\cite{LHCb:2020gnv}           \\
  $2^3S_1$     & $D_{s1}^*(2700)$     &$2922$   &$-201$     &$2722$   &$2730$   &$2714\pm5$                      \\
  $1^3P_0$     & $D_{s0}^*(2317)$     &$2540$   &$-223$     &$2316$   &$2480$   &$2317.8\pm0.5$      \\
  $1^1P_1$     & $D_{s1}(2536)$       &$2773$   &$-269$     &$2504$   &$2530$   &$2535.11\pm0.06$       \\ 
  $1^3P_1$     & $D_{s1}(2460)$       &$2700$   &$-244$     &$2456$   &$2570$   &$2459.5\pm0.6$     \\
  $1^3P_2$     & $D_{s2}^*(2573)$     &$2847$   &$-278$     &$2569$   &$2590$   &$2569.1\pm0.8$     \\
  $2^3P_0$     &                      &$3075$   &$-175$     &$2899$   &       &        \\
  $2^1P_1$     &$D_{sJ}^*(3040)$      &$3221$   &$-151$     &$3069$   &       &$3044\pm 8^{+30}_{-5}$   \\
  $2^3P_1$     &                      &$3166$   &$-187$     &$2979$   &       &              \\
  $2^3P_2$     &                      &$3278$   &$-153$     &$3134$   &       &        \\
  $1^3D_1$     & $D_{s1}^*(2860)$     &$3030$   &$-184$      &$2846$   &$2900$   &$2859\pm27$    \\
  $1^1D_2$     &                      &$3112$   &$-253$      &$2858$   &       &       \\
  $1^3D_2$     &                      &$3092$   &$-239$      &$2853$   &       &      \\
  $1^3D_3$     & $D_{s3}^*(2860)$     &$3154$   &$-286$      &$2868$   &$2920$   &$2860\pm7$       \\
  \hline\hline

\end{tabular}
\end{center}
\end{table*}

\begin{table*}
\caption{\label{tab:shift} Mass shift (in MeV) from the coupled-channels. The
coupling of the quark core to the molecular component is $\gamma_0=0.478$. } 
\begin{tabular}{ccccccccccccc} 
\hline 
\hline \\
State                & $DK$  & $DK^*$ & $D^*K$ & $D^*K^*$ & $D_s\eta$  & $D_s\eta^\prime$  & $D_s\phi$  & $D_s^*\eta$ &$D_s^*\eta^\prime$  &$D_s^*\phi$ & Total \\\hline
\hline \\
$1^1S_0$             &$0$        &$-39$    &$-44$    &$-71$    &$0$   &$0$   &$-10$    &$-8$    &$-3$   &$-19$   &$-195$  \\  
$1^3S_1$             &$-20$      &$-30$    &$-34$    &$-92$    &$-3$  &$-1$  &$-8$     &$-6$    &$-2$   &$-24$   &$-221$  \\ 
$2^1S_0$             &$0$        &$-46$    &$-62$    &$-73$    &$0$   &$0$   &$-8$     &$-8$    &$-2$   &$-14$   &$-213$  \\ 
$2^3S_1$             &$-2$       &$-38$    &$-28$    &$-96$    &$-3$  &$-1$  &$-6$     &$-7$    &$-2$   &$-17$   &$-201$  \\   
$1^3P_0$             &$-66$      &$0$      &$0$      &$-122$   &$-5$  &$-2$  &$0$      &$0$     &$0$    &$-28$   &$-223$  \\ 
$1^1P_1$             &$0$        &$-52$    &$-89$    &$-87$    &$0$   &$0$   &$-10$    &$-9$    &$-3$   &$-19$   &$-269$  \\      
$1^3P_1$             &$0$        &$-40$    &$-66$    &$-98$    &$0$   &$0$   &$-8$    &$-7$    &$-2$   &$-22$   &$-244$  \\ 
$1^3P_2$             &$-44$        &$-39$    &$-50$    &$-100$    &$-6$   &$-2$   &$-8$    &$-7$    &$-3$   &$-20$   &$-278$  \\ 

$2^3P_0$             &$-19$        &$0$    &$0$    &$-133$    &$-3$   &$-3$   &$0$    &$0$    &$0$   &$-18$   &$-175$  \\ 
$2^1P_1$             &$0$        &$-30$    &$-13$    &$-79$    &$0$   &$0$   &$-8$    &$-4$    &$-3$   &$-15$   &$-151$  \\    
$2^3P_1$             &$0$        &$-29$    &$-23$    &$-107$    &$0$   &$0$   &$-10$    &$-3$    &$-2$   &$-14$   &$-187$  \\  
$2^3P_2$             &$-10$        &$-11$    &$-8$    &$-80$    &$-1$   &$-1$   &$-7$    &$-2$    &$-2$   &$-31$   &$-153$  \\    

$1^3D_1$             &$13$        &$-19$    &$0$    &$-146$    &$-1$   &$-1$   &$-2$    &$-1$    &$0$   &$-26$   &$-184$  \\  
$1^1D_2$             &$0$        &$-66$    &$-53$    &$-98$    &$0$   &$0$   &$-9$    &$-8$    &$-3$   &$-16$   &$-253$  \\ 
$1^3D_2$             &$0$        &$-62$    &$-35$    &$-106$    &$0$   &$0$   &$-8$    &$-7$    &$-2$   &$-19$   &$-239$  \\  
$1^3D_3$             &$-44$        &$-43$    &$-52$    &$-109$    &$-6$   &$-2$   &$-7$    &$-6$    &$-2$   &$-15$   &$-286$  \\ 
\hline 
\hline
\end{tabular}
\end{table*}

\begin{table*}[htpb]
\begin{center}
\caption{ \label{tab:dsdeday1}The strong decay widths (in MeV) of charmed-strange mesons. The symbol `$-$' in the table means the decay mode is forbidden or there is no experimental information.}
\footnotesize
\begin{tabular}{cccccccccccccc}
\hline\hline
  $n^{2S+1}L_J$  & states         & PDG~\cite{ParticleDataGroup:2022pth} & $DK$  & $DK^*$ & $D^*K$ & $D^*K^*$ & $D_s\eta$  & $D_s\eta^\prime$  & $D_s\phi$  & $D_s^*\eta$ &$D_s\eta^\prime$  &$D_s^*\phi$ & Total \\\hline
  $2^1S_0$   & $D_{s0}(2590)$     &$89\pm16\pm12$\cite{LHCb:2020gnv}    &$-$      & $-$     & 87   & $-$   & $-$  & $-$  &$-$  &$-$  &$-$  &$-$   &$87$ \\
  $2^3S_1$     & $D_{s1}^*(2700)$ &$122\pm10$       & $32$    &$-$      & $77$    &$-$     & 6    &$-$   &$-$   & 3   &$-$  &$-$  & $119$ \\
  $1^3P_0$     & $D_{s0}^*(2317)$ &$<3.8$           &$-$      &$-$     &$-$      &$-$     &$-$   &$-$   &$-$   &$-$  &$-$  &$-$  &$-$     \\
  $1^1P_1$     & $D_{s1}(2536)$  &$0.92\pm0.05$    &$-$      &$-$      & $10$    &$-$     &$-$   &$-$   &$-$   &$-$  &$-$  &$-$  & $10$  \\
  $1^3P_1$     & $D_{s1}(2460)$   &$<3.5$           &$-$      &$-$      &$-$      &$-$     &$-$   &$-$   &$-$   &$-$  & $-$  &$-$  &$-$ \\
  $1^3P_2$     & $D_{s2}^*(2573)$ &$16.9\pm0.7$     & $15$    &$-$      & $1$     &$-$     &$-$   &$-$   &$-$   &$-$  &$-$  &$-$  & $17$  \\
  $1^3D_1$     & $D_{s1}^*(2860)$ &$159\pm80$       & $46$    & $27$    & $35$    &$-$     & 7   &$-$    &$-$   & 3   &$-$  &$-$  & $118$  \\ 
  $1^1D_2$     &  $-$             &$-$              &$-$      &38       &62       &$-$     &$-$   &$-$     &$-$    &4     &$-$   &$-$   &104    \\
  $1^3D_2$     &  $-$             &$-$              &$-$      & 53      & 75      &$-$       &$-$   &$-$      &$-$     & 6    &$-$     &$-$      &134    \\
  $1^3D_3$     & $D_{s3}^*(2860)$ &$53\pm10$        & $39$    & $2$     & $22$    & $-$     & 2   &$-$   &$-$   &$-$  &$-$  &$-$  & $65$  \\
  \hline\hline

\end{tabular}
\end{center}
\end{table*}

\begin{table}[htpb]
\begin{center}
\caption{\label{tab:dsdecay2} The decay width~(in MeV) of $2P$ charmed-strange mesons. The symbol `$-$' in the table means the decay mode is forbidden or there is no experimental information.}
\footnotesize
\begin{tabular}{lccccccccc}
\hline\hline

  Channel               &$2^3P_0$   &$2^1P_1$              &$2^3P_1$   &$2^3P_2$    \\   
                        &$-$        & $D_{sJ}^*(3040)$   &$-$           &$-$         \\\hline
  $DK$                  &52         &$-$                 &$-$           &1           \\
  $DK^*$                &$-$        &67                  &25            &45             \\
  $D^*K$                &$-$        &50                  &50             &17          \\
  $D^*K^*$              &3          &77                  &29           &93        \\
  $D_s\eta$             &1          &$-$                 &$-$           &3           \\
  $D_s\eta^\prime$      &$-$        &$-$                 &$-$           &2          \\
  $D_s\phi$             &$-$        &7                   &$-$            &6           \\
  $D_s^*\eta$           &$-$        &8                   &3             &6          \\
  $D_s^*\eta^\prime$    &$-$        &$-$                   &$-$             &$-$            \\
  $D_s^*\phi$           &$-$        &$-$                &$-$            &12          \\
  $DK^*_0(1430)$        &$-$        &$-$                &$-$           &$-$           \\
  $DK_{1B}$             &$-$        &$-$                &$-$             &1           \\
  $DK_{1A}$             &$-$        &$-$                &$-$           &$-$            \\
  $DK^*_2(1430)$        &$-$        &$-$                &$-$           &$-$          \\
  $D_0^*(2300)K$        &$-$        &$-$                &$-$           &$-$         \\
  $D_1(2420)K$          &$-$        &$-$                &12             &8            \\
  $D_1(2430)K$          &$-$        &$-$                &1          &3            \\
  $D_2^*(2460)K$        &$-$        &38                 &5           &18            \\
  Total                 &57         &247                &127           &215       \\
  Exp.                  & $-$       &$239\pm35^{+46}_{-42}$         &$-$           & $-$       \\
  \hline\hline

\end{tabular}
\end{center}
\end{table}

\begin{table*}
\caption{\label{tab:probabilities} Probabilities ($\%$) of every coupled-channel component and bare $c\Bar{s}$ component for the charmed-strange mesons below threshold. } 
\begin{tabular}{ccccccccccccccc} 
\hline 
\hline \\
 State      &     & $DK$  & $DK^*$ & $D^*K$ & $D^*K^*$ & $D_s\eta$  & $D_s\eta^\prime$  & $D_s\phi$  & $D_s^*\eta$ &$D_s^*\eta^\prime$  &$D_s^*\phi$ & $P_{molecule}$   &$P_{c\bar{s}}$ \\\hline
\hline \\
$1^1S_0$    &$D_s$            &$0$   &$3$   &$4$   &$4$   &$0$   &$0$   &$1$   &$1$   &$0$   &$1$   &$15$   &$85$ \\  
$1^3S_1$    &$D_{s}^{*}$      &$2$   &$2$   &$3$   &$6$   &$0$   &$0$   &$0$   &$0$   &$0$   &$1$   &$17$   &$83$ \\  
$1^3P_0$    &$D_{s0}^*(2317)$ &$29$   &$0$   &$0$   &$7$   &$1$   &$0$   &$0$   &$0$   &$0$   &$1$   &$38$   &$62$ \\  
$1^3P_1$    &$D_{s1}(2460)$   &$0$   &$4$   &$23$   &$6$   &$0$   &$0$   &$0$   &$1$   &$0$   &$1$   &$36$   &$64$ \\  
\hline 
\hline
\end{tabular}
\end{table*}

In this section, we present our numerical calculation results of the mass spectrum and the strong decay widths for the charmed-strange mesons.
As we known, the Godfrey-Isgur model is a quenched quark model, and  the unquenched effects are already absorbed into the model parameters. Since we consider the unquenched coupled-channel effects explicitly in this work, the parameters of the Godfrey-Isgur model should be  different from the ones of Ref.~\cite{Godfrey:1985xj}, which will give rise to the different bare masses.
Firstly, the free parameters of the GI model are obtained by fitting to the masses of the charmed-strange mesons, which are listed in Table~\ref{giparameter}. In our fitting, the input states include the $D_s$ $(1^1S_0)$, $D_s^*$ $(1^3S_1)$, $D_{s0}^*(2317)$ $(1^3P_0)$, $D_{s2}^*(2573)$ $(1^3P_2)$, $D_{s1}^*(2860)$ $(1^3D_1)$, $D_{s3}^*(2860)$ $(1^3D_3)$, $D_{s1}^*(2700)$ $(2^3S_1)$ and $D_{s0}(2590)$ $(2^1S_0)$. Then, the physical masses and the corresponding mass shifts of the mesons could be simultaneously determined, which are shown in Table~\ref{tab:dsm}. We also present the mass shifts of every channel in Table~\ref{tab:shift}\footnote{
According Table~\ref{tab:shift}, the $D^*K^*$ contribution to the mass shift is large for all the states. The mass shifts of the coupled-channels are stable up to an overall multiplier $\gamma^2$, which is obtained by fitted to the width of $D_{s2}^*(2573)$ here. Indeed, it is maybe due to the spin-enhancement effect.}. The strong decay widths of the charmed-strange mesons are given in Table~\ref{tab:dsdeday1} and Table \ref{tab:dsdecay2}. In addition, the probabilities of every coupled-channel and $c\bar{s}$ component for the states below the threshold are listed in Table~\ref{tab:probabilities}.

Taking into account the coupled-channel effects, the masses of the ground states $D_s$ and $D_s^*$ are determined to be 1968~MeV and 2112~MeV, which are in good agreement with the experimental data. According to Table~\ref{tab:probabilities}, the probabilities of the $c\bar{s}$ component are 85\% and 83\% for the  $D_s$ and $D_s^*$, respectively, which implies that the $c\bar{s}$ component is dominant.

According to Eqs.~(\ref{deltam}) and (\ref{eqn:Tdag}), the mass shifts $\Delta M$ are related to the parameter $\gamma$ of the $^3P_0$ model, which  represents the probability that a quark-antiquark pair is created from the vacuum. However, the probabilities of the molecular components are determined by the derivative of the mass shifts, not the mass shifts, which could be easily understood as follows,
\begin{eqnarray}
	P_{BC} &= & \left[1+\sum_{BC\ell J} \int_0^{\infty} p^2 dp \mbox{ } \frac{\left|\left\langle BC;p \right| T^\dag \left| A \right\rangle \right|^2}{(M - E_{BC})^2}\right]^{-1}\nonumber  \\
& & \times \int_0^{\infty} p^2 dp \mbox{ } \frac{\left|\left\langle BC;p \right| T^\dag \left| A \right\rangle \right|^2}{(M - E_{BC})^2} \nonumber \\
& \propto  &  \frac{\partial \Delta M_{BC}}{\partial M},\\
\Delta M_{BC} &=& \int_0^{\infty} p^2 dp \mbox{ } \frac{\left|\left\langle BC;p \right| T^\dag \left| A \right\rangle \right|^2}{M - E_{BC}+i\epsilon}.
\end{eqnarray}
Indeed, it is shown that the mass shifts account for $10\sim 20\%$ of the quenched mass values in Ref.~\cite{Chen:2017mug}, which is consistent with our predictions.

The masses of the $D_{s}(1^3P_0)$ and $D_{s1}(1^3P_1)$ states are numerically estimated to be 2316~MeV and 2456~MeV, which are in good agreement with the experimental data of the $D_{s0}^*(2317)$ and $D_{s1}(2460)$, both of which have large non-$c\bar{s}$ component. 
The biggest non-$c\bar{s}$ component of $D_{s0}^*(2317)$ is $DK$, and the one of $D_{s1}(2460)$ is $D^*K$, consistently with the phenomenological results of Ref.~\cite{Ortega:2016mms}. 
It should be noted that, the predicted probabilities of the  molecular components $DK$ and $D^*K$  for the $D^*_{s0}(2317)$ and $D_{s1}(2460)$ are smaller than those obtained from a lattice
analysis of these resonances in Ref.~\cite{MartinezTorres:2014kpc}, which is maybe due to the neglect of the explicit meson-meson interactions in this work, and the consideration of the explicit meson-meson interactions could bring the results of the present work in closer agreement with these lattice data. These is indeed the case, in general, as shown in Refs.~\cite{Song:2023pdq,Dai:2023kwv}.

Our predictions for the mass and width of $D_{s}(2^3S_1)$ are 2722~MeV and 119~MeV, respectively, which are in good agreement with the experimental data for the $D_{s1}(2700)$ mass ($2714\pm5$~MeV) and width ($112\pm10$~MeV). The dominant decay modes are $DK$ and $D^*K$, consistently with
the experimental measurements~\cite{ParticleDataGroup:2022pth}. Our results support the interpretation of $D_{s1}(2700)$  as the $D_{s}(2^3S_1)$ state.

We predict that the mass and width of $D_{s}(1^1P_1)$ are 2504~MeV and 10~MeV, close to the experimental data of the $D_{s1}(2536)$, and the dominant decay mode is $D^*K$, consistently with the experimental measurements of $D_{s1}(2536)$~\cite{ParticleDataGroup:2022pth},  which implies that $D_{s1}(2536)$ could be well interpreted as the $D_{s}(1^1P_1)$ state. For the $D_{s}(1^3P_2)$, the predicted mass and width are 2569~MeV and 17~MeV, in good agreement with the experimental data of the $D^*_{s2}(2573)$, and its dominant decay mode is $DK$,  consistently with the experimental measurements of $D^*_{s2}(2573)$~\cite{ParticleDataGroup:2022pth}, which supports that the state could be well interpreted as the $D_{s}(1^3P_2)$ state.

The mass and width of the  $D_s(1^3D_1)$ are predicted to be 2846~MeV and 118~MeV, which are in good agreement with the experimentally measured $D^*_{s1}(2860)$ mass $2859\pm27$~MeV and width $159\pm 80$~MeV. In addition, the mass and width of the  $D_s(1^3D_3)$ are calculated to be 2868~MeV and 65~MeV, which are in good agreement with the $D^*_{s3}(2860)$ mass $2860\pm7$~MeV and width $53\pm 10$~MeV. The dominant decay mode of $D_s(1^3D_1)$ and $D_s(1^3D_3)$ is predicted to be $DK$, consistently with the experimental measurements~\cite{ParticleDataGroup:2022pth}.  Our results suggest that  $D^*_{s1}(2860)$ and $D^*_{s3}(2860)$ can be interpreted as the $D_s(1^3D_1)$ and $D_s(1^3D_3)$, respectively, supported by Refs.~\cite{Godfrey:2014fga,Song:2014mha,Song:2015nia}.

The $D_{sJ}(3040)$ was observed in $D^*K$ mass spectrum by  BaBar Collaboration~\cite{BaBar:2009rro}, and its mass and width are determined to be $3044\pm 8^{+30}_{-5}$~MeV and $239\pm 35^{+46}_{-42}$~MeV, respectively. Its mass is consistent with the predicted mass (3069~MeV) of the $D_s(2^1P_1)$. We have calculated its strong decay width by regarding $D_{sJ}(3040)$ as $D_s(2^1P_1)$, which is 247~MeV, in good agreement with the experimental value. Thus, our results support that $D_{sJ}(3040)$ could be well interpreted as the $D_s(2^1P_1)$ state.

The recently observed state $D_{s0}(2590)$ has the spin-parity quantum numbers $J^P=1^-$, and is suggested to be the candidate of $D_{s}(2^1S_0)$ by LHCb~\cite{LHCb:2020gnv}. One can see that the predicted  mass of $D_{s}(2^1S_0)$ in Table \ref{tab:dsm} is 2646~MeV, a little higher than the experimental data, but the predicted width 87~MeV is in good agreement with the experimental value.
Indeed, in Ref.~\cite{Godfrey:1985xj}, the predicted masses of $D_{s}(2^1S_0)$ are higher than $D_{s0}(2590)$. Thus, our results suggest that $D_{s0}(2590)$ could be the candidate of the $D_{s}(2^1S_0)$. Taking into account that $D_{s0}(2590)$ was only observed by LHCb, we strongly encourage the experimental side to search for this state in other processes, and the more precise information about the $D_{s0}(2590)$ could shed light on its assignment.

In addition, we have predicted the masses and the strong decay properties of the charmed-strange mesons around 3~GeV. For the $D$-wave states, the masses of the $D_s(1^1D_2)$ and $D_s(1^3D_2)$ are predicted to be 2858~MeV and 2853~MeV, respectively, and their widths are predicted to be 104~MeV and 134~MeV, respectively. Their dominant decay modes are $D^*K$ and $DK^*$. For the $2P$ states, the masses of the $D_s(2^3P_0)$, $D_s(2^3P_1)$, and $D_s(2^3P_2)$ are predicted to be 2899~MeV, 2979~MeV, and 3134~MeV, respectively, and their widths are predicted to be 57~MeV, 127~MeV, and 215~MeV. The dominant decay mode of $D_s(2^3P_0)$ is $DK$, while the ones of $D_s(2^3P_1)$ and $D_s(2^3P_2)$ are $DK^*$, $D^*K$, and $D^*K^*$. Our predictions would be helpful to search for these states experimentally.

\section{Summary}
\label{sec:summary}
In this work, we have investigated the mass spectrum and the strong decay properties of the charmed-strange mesons within the Godfrey-Isgur model by considering the coupled-channel effects.  The bare mass is obtained by the Godfrey-Isgur model, and the mass shift from the coupled-channel effects is given by the interaction between the initial meson and the coupled channels, with the interaction being described by the quark-antiquark creation operator from the $^3P_0$ model. One find that the  $D^*K^*$ contribution to the mass shift is large for all the charmed-strange states, which is maybe due to the spin-enhancement effect.

Our results show that $D_{s0}^*(2317)$ and $D_{s1}(2460)$ can be interpreted as the $D_{s}(1^3P_0)$ and $D_{s}(1^3P_1)$ states with larger $DK$ and $D^*K$ components, respectively, although the probabilities of the $DK$ and $D^*K$ components for $D_{s0}^*(2317)$ and $D_{s1}(2460)$ are smaller than other results, which may be due to our neglect of the direct interaction of the meson components.
Comparing our theoretical predicted results with the experimental measurement, it is found that $D_{s1}(2700)$, $D_{s1}(2536)$, $D^*_{s2}(2573)$, $D_{s1}^*(2860)$, $D_{s3}^*(2860)$, and $D_{sJ}^*(3040)$ could be well interpreted as the $D_s(2^3S_1)$, $D_s(1^1P_1)$, $D_s(1^3P_2)$, $D_s(1^3D_1)$, $D_s(1^3D_3)$, and $D_s(2^1P_1)$ states, respectively. 

For the recently observed state $D_{s0}(2590)$,
although its mass is about 50~MeV less than our prediction for the $D_{s}(2^1S_0)$ state, its width is still in good agreement with that of $D_{s}(2^1S_0)$, which implies that $D_{s0}(2590)$ could be assigned as the candidate of $D_{s}(2^1S_0)$.  We emphasize that $D_{s0}(2590)$ needs to be further confirmed by the experimental measurement, and the more precise information about $D_{s0}(2590)$ would shed light on the assignment of $D_{s}(2^1S_0)$.
Furthermore, we have predicted the masses and the strong decay properties of the charmed-strange mesons with masses around 3~GeV, which would be helpful to experimentally search for these states.  
 
\section{Acknowledgements}
We warmly thank Eulogio Oset for careful reading and useful comments.
This work is partly supported by the National Natural Science Foundation of China under Grants Nos. 12192263, 12005191, and 11905187,
the Natural Science Foundation of Henan under Grant Nos. 222300420554, 232300421140, the Project of Youth Backbone Teachers of Colleges and Universities of Henan Province (2020GGJS017), and the Open Project of Guangxi Key Laboratory of Nuclear Physics and Nuclear Technology, No. NLK2021-08.
.

\bibliography{cite}  

\begin{thebibliography}{69}
\expandafter\ifx\csname natexlab\endcsname\relax\def\natexlab#1{#1}\fi
\expandafter\ifx\csname bibnamefont\endcsname\relax
  \def\bibnamefont#1{#1}\fi
\expandafter\ifx\csname bibfnamefont\endcsname\relax
  \def\bibfnamefont#1{#1}\fi
\expandafter\ifx\csname citenamefont\endcsname\relax
  \def\citenamefont#1{#1}\fi
\expandafter\ifx\csname url\endcsname\relax
  \def\url#1{\texttt{#1}}\fi
\expandafter\ifx\csname urlprefix\endcsname\relax\def\urlprefix{URL }\fi
\providecommand{\bibinfo}[2]{#2}
\providecommand{\eprint}[2][]{\url{#2}}

\bibitem[{\citenamefont{Aaij et~al.}(2021)}]{LHCb:2020gnv}
\bibinfo{author}{\bibfnamefont{R.}~\bibnamefont{Aaij}} \bibnamefont{et~al.}
  (\bibinfo{collaboration}{LHCb}), \bibinfo{journal}{Phys. Rev. Lett.}
  \textbf{\bibinfo{volume}{126}}, \bibinfo{pages}{122002}
  (\bibinfo{year}{2021}), \eprint{2011.09112}.

\bibitem[{\citenamefont{Workman et~al.}(2022)}]{ParticleDataGroup:2022pth}
\bibinfo{author}{\bibfnamefont{R.~L.} \bibnamefont{Workman}}
  \bibnamefont{et~al.} (\bibinfo{collaboration}{Particle Data Group}),
  \bibinfo{journal}{PTEP} \textbf{\bibinfo{volume}{2022}},
  \bibinfo{pages}{083C01} (\bibinfo{year}{2022}).

\bibitem[{\citenamefont{Rosner}(2007)}]{Rosner_2007}
\bibinfo{author}{\bibfnamefont{J.~L.} \bibnamefont{Rosner}},
  \bibinfo{journal}{Journal of Physics G: Nuclear and Particle Physics}
  \textbf{\bibinfo{volume}{34}}, \bibinfo{pages}{S127} (\bibinfo{year}{2007}),
  \urlprefix\url{https://doi.org/10.1088/0954-3899/34/7/s07}.

\bibitem[{\citenamefont{Segovia et~al.}(2015)\citenamefont{Segovia, Entem, and
  Fernandez}}]{Segovia:2015dia}
\bibinfo{author}{\bibfnamefont{J.}~\bibnamefont{Segovia}},
  \bibinfo{author}{\bibfnamefont{D.~R.} \bibnamefont{Entem}}, \bibnamefont{and}
  \bibinfo{author}{\bibfnamefont{F.}~\bibnamefont{Fernandez}},
  \bibinfo{journal}{Phys. Rev. D} \textbf{\bibinfo{volume}{91}},
  \bibinfo{pages}{094020} (\bibinfo{year}{2015}), \eprint{1502.03827}.

\bibitem[{\citenamefont{Chen et~al.}(2017)\citenamefont{Chen, Chen, Liu, Liu,
  and Zhu}}]{Chen:2016spr}
\bibinfo{author}{\bibfnamefont{H.-X.} \bibnamefont{Chen}},
  \bibinfo{author}{\bibfnamefont{W.}~\bibnamefont{Chen}},
  \bibinfo{author}{\bibfnamefont{X.}~\bibnamefont{Liu}},
  \bibinfo{author}{\bibfnamefont{Y.-R.} \bibnamefont{Liu}}, \bibnamefont{and}
  \bibinfo{author}{\bibfnamefont{S.-L.} \bibnamefont{Zhu}},
  \bibinfo{journal}{Rept. Prog. Phys.} \textbf{\bibinfo{volume}{80}},
  \bibinfo{pages}{076201} (\bibinfo{year}{2017}), \eprint{1609.08928}.

\bibitem[{\citenamefont{Song et~al.}(2015{\natexlab{a}})\citenamefont{Song,
  Chen, Liu, and Matsuki}}]{Song:2015nia}
\bibinfo{author}{\bibfnamefont{Q.-T.} \bibnamefont{Song}},
  \bibinfo{author}{\bibfnamefont{D.-Y.} \bibnamefont{Chen}},
  \bibinfo{author}{\bibfnamefont{X.}~\bibnamefont{Liu}}, \bibnamefont{and}
  \bibinfo{author}{\bibfnamefont{T.}~\bibnamefont{Matsuki}},
  \bibinfo{journal}{Phys. Rev. D} \textbf{\bibinfo{volume}{91}},
  \bibinfo{pages}{054031} (\bibinfo{year}{2015}{\natexlab{a}}),
  \eprint{1501.03575}.

\bibitem[{\citenamefont{Song et~al.}(2015{\natexlab{b}})\citenamefont{Song,
  Chen, Liu, and Matsuki}}]{Song:2014mha}
\bibinfo{author}{\bibfnamefont{Q.-T.} \bibnamefont{Song}},
  \bibinfo{author}{\bibfnamefont{D.-Y.} \bibnamefont{Chen}},
  \bibinfo{author}{\bibfnamefont{X.}~\bibnamefont{Liu}}, \bibnamefont{and}
  \bibinfo{author}{\bibfnamefont{T.}~\bibnamefont{Matsuki}},
  \bibinfo{journal}{Eur. Phys. J. C} \textbf{\bibinfo{volume}{75}},
  \bibinfo{pages}{30} (\bibinfo{year}{2015}{\natexlab{b}}), \eprint{1408.0471}.

\bibitem[{\citenamefont{Li and Ma}(2010)}]{Li:2009qu}
\bibinfo{author}{\bibfnamefont{D.-M.} \bibnamefont{Li}} \bibnamefont{and}
  \bibinfo{author}{\bibfnamefont{B.}~\bibnamefont{Ma}}, \bibinfo{journal}{Phys.
  Rev. D} \textbf{\bibinfo{volume}{81}}, \bibinfo{pages}{014021}
  (\bibinfo{year}{2010}), \eprint{0911.2906}.

\bibitem[{\citenamefont{Li et~al.}(2007)\citenamefont{Li, Ma, and
  Liu}}]{Li:2007px}
\bibinfo{author}{\bibfnamefont{D.-M.} \bibnamefont{Li}},
  \bibinfo{author}{\bibfnamefont{B.}~\bibnamefont{Ma}}, \bibnamefont{and}
  \bibinfo{author}{\bibfnamefont{Y.-H.} \bibnamefont{Liu}},
  \bibinfo{journal}{Eur. Phys. J. C} \textbf{\bibinfo{volume}{51}},
  \bibinfo{pages}{359} (\bibinfo{year}{2007}), \eprint{hep-ph/0703278}.

\bibitem[{\citenamefont{Ni et~al.}(2022)\citenamefont{Ni, Li, and
  Zhong}}]{Ni:2021pce}
\bibinfo{author}{\bibfnamefont{R.-H.} \bibnamefont{Ni}},
  \bibinfo{author}{\bibfnamefont{Q.}~\bibnamefont{Li}}, \bibnamefont{and}
  \bibinfo{author}{\bibfnamefont{X.-H.} \bibnamefont{Zhong}},
  \bibinfo{journal}{Phys. Rev. D} \textbf{\bibinfo{volume}{105}},
  \bibinfo{pages}{056006} (\bibinfo{year}{2022}), \eprint{2110.05024}.

\bibitem[{\citenamefont{Ebert et~al.}(2010)\citenamefont{Ebert, Faustov, and
  Galkin}}]{Ebert:2009ua}
\bibinfo{author}{\bibfnamefont{D.}~\bibnamefont{Ebert}},
  \bibinfo{author}{\bibfnamefont{R.~N.} \bibnamefont{Faustov}},
  \bibnamefont{and} \bibinfo{author}{\bibfnamefont{V.~O.}
  \bibnamefont{Galkin}}, \bibinfo{journal}{Eur. Phys. J. C}
  \textbf{\bibinfo{volume}{66}}, \bibinfo{pages}{197} (\bibinfo{year}{2010}),
  \eprint{0910.5612}.

\bibitem[{\citenamefont{Zeng et~al.}(1995)\citenamefont{Zeng, Van~Orden, and
  Roberts}}]{Zeng:1994vj}
\bibinfo{author}{\bibfnamefont{J.}~\bibnamefont{Zeng}},
  \bibinfo{author}{\bibfnamefont{J.~W.} \bibnamefont{Van~Orden}},
  \bibnamefont{and} \bibinfo{author}{\bibfnamefont{W.}~\bibnamefont{Roberts}},
  \bibinfo{journal}{Phys. Rev. D} \textbf{\bibinfo{volume}{52}},
  \bibinfo{pages}{5229} (\bibinfo{year}{1995}), \eprint{hep-ph/9412269}.

\bibitem[{\citenamefont{Godfrey and Moats}(2016)}]{Godfrey:2015dva}
\bibinfo{author}{\bibfnamefont{S.}~\bibnamefont{Godfrey}} \bibnamefont{and}
  \bibinfo{author}{\bibfnamefont{K.}~\bibnamefont{Moats}},
  \bibinfo{journal}{Phys. Rev. D} \textbf{\bibinfo{volume}{93}},
  \bibinfo{pages}{034035} (\bibinfo{year}{2016}), \eprint{1510.08305}.

\bibitem[{\citenamefont{Godfrey and Isgur}(1985)}]{Godfrey:1985xj}
\bibinfo{author}{\bibfnamefont{S.}~\bibnamefont{Godfrey}} \bibnamefont{and}
  \bibinfo{author}{\bibfnamefont{N.}~\bibnamefont{Isgur}},
  \bibinfo{journal}{Phys. Rev. D} \textbf{\bibinfo{volume}{32}},
  \bibinfo{pages}{189} (\bibinfo{year}{1985}).

\bibitem[{\citenamefont{Wang et~al.}(2022)\citenamefont{Wang, Li, Feng, Wang,
  and Liu}}]{Wang:2021orp}
\bibinfo{author}{\bibfnamefont{G.-L.} \bibnamefont{Wang}},
  \bibinfo{author}{\bibfnamefont{W.}~\bibnamefont{Li}},
  \bibinfo{author}{\bibfnamefont{T.-F.} \bibnamefont{Feng}},
  \bibinfo{author}{\bibfnamefont{Y.-L.} \bibnamefont{Wang}}, \bibnamefont{and}
  \bibinfo{author}{\bibfnamefont{Y.-B.} \bibnamefont{Liu}},
  \bibinfo{journal}{Eur. Phys. J. C} \textbf{\bibinfo{volume}{82}},
  \bibinfo{pages}{267} (\bibinfo{year}{2022}), \eprint{2107.01751}.

\bibitem[{\citenamefont{Xie et~al.}(2021)\citenamefont{Xie, Liu, and
  Geng}}]{Xie:2021dwe}
\bibinfo{author}{\bibfnamefont{J.-M.} \bibnamefont{Xie}},
  \bibinfo{author}{\bibfnamefont{M.-Z.} \bibnamefont{Liu}}, \bibnamefont{and}
  \bibinfo{author}{\bibfnamefont{L.-S.} \bibnamefont{Geng}},
  \bibinfo{journal}{Phys. Rev. D} \textbf{\bibinfo{volume}{104}},
  \bibinfo{pages}{094051} (\bibinfo{year}{2021}), \eprint{2108.12993}.

\bibitem[{\citenamefont{Ortega et~al.}(2022)\citenamefont{Ortega, Segovia,
  Entem, and Fernandez}}]{Ortega:2021fem}
\bibinfo{author}{\bibfnamefont{P.~G.} \bibnamefont{Ortega}},
  \bibinfo{author}{\bibfnamefont{J.}~\bibnamefont{Segovia}},
  \bibinfo{author}{\bibfnamefont{D.~R.} \bibnamefont{Entem}}, \bibnamefont{and}
  \bibinfo{author}{\bibfnamefont{F.}~\bibnamefont{Fernandez}},
  \bibinfo{journal}{Phys. Lett. B} \textbf{\bibinfo{volume}{827}},
  \bibinfo{pages}{136998} (\bibinfo{year}{2022}), \eprint{2111.00023}.

\bibitem[{\citenamefont{Gao et~al.}(2022)\citenamefont{Gao, Wang, L\"u, Zhu,
  and Zhao}}]{Gao:2022bsb}
\bibinfo{author}{\bibfnamefont{Z.}~\bibnamefont{Gao}},
  \bibinfo{author}{\bibfnamefont{G.-Y.} \bibnamefont{Wang}},
  \bibinfo{author}{\bibfnamefont{Q.-F.} \bibnamefont{L\"u}},
  \bibinfo{author}{\bibfnamefont{J.}~\bibnamefont{Zhu}}, \bibnamefont{and}
  \bibinfo{author}{\bibfnamefont{G.-F.} \bibnamefont{Zhao}},
  \bibinfo{journal}{Phys. Rev. D} \textbf{\bibinfo{volume}{105}},
  \bibinfo{pages}{074037} (\bibinfo{year}{2022}), \eprint{2201.00552}.

\bibitem[{\citenamefont{Hao et~al.}(2022)\citenamefont{Hao, Lu, and
  Zou}}]{Hao:2022vwt}
\bibinfo{author}{\bibfnamefont{W.}~\bibnamefont{Hao}},
  \bibinfo{author}{\bibfnamefont{Y.}~\bibnamefont{Lu}}, \bibnamefont{and}
  \bibinfo{author}{\bibfnamefont{B.-S.} \bibnamefont{Zou}},
  \bibinfo{journal}{Phys. Rev. D} \textbf{\bibinfo{volume}{106}},
  \bibinfo{pages}{074014} (\bibinfo{year}{2022}), \eprint{2208.10915}.

\bibitem[{\citenamefont{Aubert et~al.}(2003)}]{BaBar:2003oey}
\bibinfo{author}{\bibfnamefont{B.}~\bibnamefont{Aubert}} \bibnamefont{et~al.}
  (\bibinfo{collaboration}{BaBar}), \bibinfo{journal}{Phys. Rev. Lett.}
  \textbf{\bibinfo{volume}{90}}, \bibinfo{pages}{242001}
  (\bibinfo{year}{2003}), \eprint{hep-ex/0304021}.

\bibitem[{\citenamefont{Besson et~al.}(2003)}]{CLEO:2003ggt}
\bibinfo{author}{\bibfnamefont{D.}~\bibnamefont{Besson}} \bibnamefont{et~al.}
  (\bibinfo{collaboration}{CLEO}), \bibinfo{journal}{Phys. Rev. D}
  \textbf{\bibinfo{volume}{68}}, \bibinfo{pages}{032002}
  (\bibinfo{year}{2003}), \bibinfo{note}{[Erratum: Phys.Rev.D 75, 119908
  (2007)]}, \eprint{hep-ex/0305100}.

\bibitem[{\citenamefont{Di~Pierro and Eichten}(2001)}]{DiPierro:2001dwf}
\bibinfo{author}{\bibfnamefont{M.}~\bibnamefont{Di~Pierro}} \bibnamefont{and}
  \bibinfo{author}{\bibfnamefont{E.}~\bibnamefont{Eichten}},
  \bibinfo{journal}{Phys. Rev. D} \textbf{\bibinfo{volume}{64}},
  \bibinfo{pages}{114004} (\bibinfo{year}{2001}), \eprint{hep-ph/0104208}.

\bibitem[{\citenamefont{Bali}(2003)}]{Bali:2003jv}
\bibinfo{author}{\bibfnamefont{G.~S.} \bibnamefont{Bali}},
  \bibinfo{journal}{Phys. Rev. D} \textbf{\bibinfo{volume}{68}},
  \bibinfo{pages}{071501} (\bibinfo{year}{2003}), \eprint{hep-ph/0305209}.

\bibitem[{\citenamefont{Dougall et~al.}(2003)\citenamefont{Dougall, Kenway,
  Maynard, and McNeile}}]{Dougall:2003hv}
\bibinfo{author}{\bibfnamefont{A.}~\bibnamefont{Dougall}},
  \bibinfo{author}{\bibfnamefont{R.~D.} \bibnamefont{Kenway}},
  \bibinfo{author}{\bibfnamefont{C.~M.} \bibnamefont{Maynard}},
  \bibnamefont{and} \bibinfo{author}{\bibfnamefont{C.}~\bibnamefont{McNeile}}
  (\bibinfo{collaboration}{UKQCD}), \bibinfo{journal}{Phys. Lett. B}
  \textbf{\bibinfo{volume}{569}}, \bibinfo{pages}{41} (\bibinfo{year}{2003}),
  \eprint{hep-lat/0307001}.

\bibitem[{\citenamefont{Guo et~al.}(2018)\citenamefont{Guo, Hanhart,
  Mei\ss{}ner, Wang, Zhao, and Zou}}]{Guo:2017jvc}
\bibinfo{author}{\bibfnamefont{F.-K.} \bibnamefont{Guo}},
  \bibinfo{author}{\bibfnamefont{C.}~\bibnamefont{Hanhart}},
  \bibinfo{author}{\bibfnamefont{U.-G.} \bibnamefont{Mei\ss{}ner}},
  \bibinfo{author}{\bibfnamefont{Q.}~\bibnamefont{Wang}},
  \bibinfo{author}{\bibfnamefont{Q.}~\bibnamefont{Zhao}}, \bibnamefont{and}
  \bibinfo{author}{\bibfnamefont{B.-S.} \bibnamefont{Zou}},
  \bibinfo{journal}{Rev. Mod. Phys.} \textbf{\bibinfo{volume}{90}},
  \bibinfo{pages}{015004} (\bibinfo{year}{2018}), \bibinfo{note}{[Erratum:
  Rev.Mod.Phys. 94, 029901 (2022)]}, \eprint{1705.00141}.

\bibitem[{\citenamefont{Chen et~al.}(2023)\citenamefont{Chen, Chen, Liu, Liu,
  and Zhu}}]{Chen:2022asf}
\bibinfo{author}{\bibfnamefont{H.-X.} \bibnamefont{Chen}},
  \bibinfo{author}{\bibfnamefont{W.}~\bibnamefont{Chen}},
  \bibinfo{author}{\bibfnamefont{X.}~\bibnamefont{Liu}},
  \bibinfo{author}{\bibfnamefont{Y.-R.} \bibnamefont{Liu}}, \bibnamefont{and}
  \bibinfo{author}{\bibfnamefont{S.-L.} \bibnamefont{Zhu}},
  \bibinfo{journal}{Rept. Prog. Phys.} \textbf{\bibinfo{volume}{86}},
  \bibinfo{pages}{026201} (\bibinfo{year}{2023}), \eprint{2204.02649}.

\bibitem[{\citenamefont{Oset et~al.}(2016)}]{Oset:2016lyh}
\bibinfo{author}{\bibfnamefont{E.}~\bibnamefont{Oset}} \bibnamefont{et~al.},
  \bibinfo{journal}{Int. J. Mod. Phys. E} \textbf{\bibinfo{volume}{25}},
  \bibinfo{pages}{1630001} (\bibinfo{year}{2016}), \eprint{1601.03972}.

\bibitem[{\citenamefont{Albaladejo et~al.}(2018)\citenamefont{Albaladejo,
  Fernandez-Soler, Nieves, and Ortega}}]{Albaladejo:2018mhb}
\bibinfo{author}{\bibfnamefont{M.}~\bibnamefont{Albaladejo}},
  \bibinfo{author}{\bibfnamefont{P.}~\bibnamefont{Fernandez-Soler}},
  \bibinfo{author}{\bibfnamefont{J.}~\bibnamefont{Nieves}}, \bibnamefont{and}
  \bibinfo{author}{\bibfnamefont{P.~G.} \bibnamefont{Ortega}},
  \bibinfo{journal}{Eur. Phys. J. C} \textbf{\bibinfo{volume}{78}},
  \bibinfo{pages}{722} (\bibinfo{year}{2018}), \eprint{1805.07104}.

\bibitem[{\citenamefont{Kolomeitsev and Lutz}(2004)}]{Kolomeitsev:2003ac}
\bibinfo{author}{\bibfnamefont{E.~E.} \bibnamefont{Kolomeitsev}}
  \bibnamefont{and} \bibinfo{author}{\bibfnamefont{M.~F.~M.}
  \bibnamefont{Lutz}}, \bibinfo{journal}{Phys. Lett. B}
  \textbf{\bibinfo{volume}{582}}, \bibinfo{pages}{39} (\bibinfo{year}{2004}),
  \eprint{hep-ph/0307133}.

\bibitem[{\citenamefont{Hofmann and Lutz}(2004)}]{Hofmann:2003je}
\bibinfo{author}{\bibfnamefont{J.}~\bibnamefont{Hofmann}} \bibnamefont{and}
  \bibinfo{author}{\bibfnamefont{M.~F.~M.} \bibnamefont{Lutz}},
  \bibinfo{journal}{Nucl. Phys. A} \textbf{\bibinfo{volume}{733}},
  \bibinfo{pages}{142} (\bibinfo{year}{2004}), \eprint{hep-ph/0308263}.

\bibitem[{\citenamefont{Barnes et~al.}(2003)\citenamefont{Barnes, Close, and
  Lipkin}}]{Barnes:2003dj}
\bibinfo{author}{\bibfnamefont{T.}~\bibnamefont{Barnes}},
  \bibinfo{author}{\bibfnamefont{F.~E.} \bibnamefont{Close}}, \bibnamefont{and}
  \bibinfo{author}{\bibfnamefont{H.~J.} \bibnamefont{Lipkin}},
  \bibinfo{journal}{Phys. Rev. D} \textbf{\bibinfo{volume}{68}},
  \bibinfo{pages}{054006} (\bibinfo{year}{2003}), \eprint{hep-ph/0305025}.

\bibitem[{\citenamefont{Chen and Li}(2004)}]{Chen:2004dy}
\bibinfo{author}{\bibfnamefont{Y.-Q.} \bibnamefont{Chen}} \bibnamefont{and}
  \bibinfo{author}{\bibfnamefont{X.-Q.} \bibnamefont{Li}},
  \bibinfo{journal}{Phys. Rev. Lett.} \textbf{\bibinfo{volume}{93}},
  \bibinfo{pages}{232001} (\bibinfo{year}{2004}), \eprint{hep-ph/0407062}.

\bibitem[{\citenamefont{Guo et~al.}(2006)\citenamefont{Guo, Shen, Chiang, Ping,
  and Zou}}]{Guo:2006fu}
\bibinfo{author}{\bibfnamefont{F.-K.} \bibnamefont{Guo}},
  \bibinfo{author}{\bibfnamefont{P.-N.} \bibnamefont{Shen}},
  \bibinfo{author}{\bibfnamefont{H.-C.} \bibnamefont{Chiang}},
  \bibinfo{author}{\bibfnamefont{R.-G.} \bibnamefont{Ping}}, \bibnamefont{and}
  \bibinfo{author}{\bibfnamefont{B.-S.} \bibnamefont{Zou}},
  \bibinfo{journal}{Phys. Lett. B} \textbf{\bibinfo{volume}{641}},
  \bibinfo{pages}{278} (\bibinfo{year}{2006}), \eprint{hep-ph/0603072}.

\bibitem[{\citenamefont{Guo et~al.}(2007)\citenamefont{Guo, Shen, and
  Chiang}}]{Guo:2006rp}
\bibinfo{author}{\bibfnamefont{F.-K.} \bibnamefont{Guo}},
  \bibinfo{author}{\bibfnamefont{P.-N.} \bibnamefont{Shen}}, \bibnamefont{and}
  \bibinfo{author}{\bibfnamefont{H.-C.} \bibnamefont{Chiang}},
  \bibinfo{journal}{Phys. Lett. B} \textbf{\bibinfo{volume}{647}},
  \bibinfo{pages}{133} (\bibinfo{year}{2007}), \eprint{hep-ph/0610008}.

\bibitem[{\citenamefont{Gamermann et~al.}(2007)\citenamefont{Gamermann, Oset,
  Strottman, and Vicente~Vacas}}]{Gamermann:2006nm}
\bibinfo{author}{\bibfnamefont{D.}~\bibnamefont{Gamermann}},
  \bibinfo{author}{\bibfnamefont{E.}~\bibnamefont{Oset}},
  \bibinfo{author}{\bibfnamefont{D.}~\bibnamefont{Strottman}},
  \bibnamefont{and} \bibinfo{author}{\bibfnamefont{M.~J.}
  \bibnamefont{Vicente~Vacas}}, \bibinfo{journal}{Phys. Rev. D}
  \textbf{\bibinfo{volume}{76}}, \bibinfo{pages}{074016}
  (\bibinfo{year}{2007}), \eprint{hep-ph/0612179}.

\bibitem[{\citenamefont{Gamermann and Oset}(2007)}]{Gamermann:2007fi}
\bibinfo{author}{\bibfnamefont{D.}~\bibnamefont{Gamermann}} \bibnamefont{and}
  \bibinfo{author}{\bibfnamefont{E.}~\bibnamefont{Oset}},
  \bibinfo{journal}{Eur. Phys. J. A} \textbf{\bibinfo{volume}{33}},
  \bibinfo{pages}{119} (\bibinfo{year}{2007}), \eprint{0704.2314}.

\bibitem[{\citenamefont{Lutz and Soyeur}(2008)}]{Lutz:2007sk}
\bibinfo{author}{\bibfnamefont{M.~F.~M.} \bibnamefont{Lutz}} \bibnamefont{and}
  \bibinfo{author}{\bibfnamefont{M.}~\bibnamefont{Soyeur}},
  \bibinfo{journal}{Nucl. Phys. A} \textbf{\bibinfo{volume}{813}},
  \bibinfo{pages}{14} (\bibinfo{year}{2008}), \eprint{0710.1545}.

\bibitem[{\citenamefont{Molina et~al.}(2010)\citenamefont{Molina, Branz, and
  Oset}}]{Molina:2010tx}
\bibinfo{author}{\bibfnamefont{R.}~\bibnamefont{Molina}},
  \bibinfo{author}{\bibfnamefont{T.}~\bibnamefont{Branz}}, \bibnamefont{and}
  \bibinfo{author}{\bibfnamefont{E.}~\bibnamefont{Oset}},
  \bibinfo{journal}{Phys. Rev. D} \textbf{\bibinfo{volume}{82}},
  \bibinfo{pages}{014010} (\bibinfo{year}{2010}), \eprint{1005.0335}.

\bibitem[{\citenamefont{Kalashnikova}(2005)}]{Kalashnikova:2005ui}
\bibinfo{author}{\bibfnamefont{Y.~S.} \bibnamefont{Kalashnikova}},
  \bibinfo{journal}{Phys. Rev. D} \textbf{\bibinfo{volume}{72}},
  \bibinfo{pages}{034010} (\bibinfo{year}{2005}), \eprint{hep-ph/0506270}.

\bibitem[{\citenamefont{Li et~al.}(2009)\citenamefont{Li, Meng, and
  Chao}}]{Li:2009ad}
\bibinfo{author}{\bibfnamefont{B.-Q.} \bibnamefont{Li}},
  \bibinfo{author}{\bibfnamefont{C.}~\bibnamefont{Meng}}, \bibnamefont{and}
  \bibinfo{author}{\bibfnamefont{K.-T.} \bibnamefont{Chao}},
  \bibinfo{journal}{Phys. Rev. D} \textbf{\bibinfo{volume}{80}},
  \bibinfo{pages}{014012} (\bibinfo{year}{2009}), \eprint{0904.4068}.

\bibitem[{\citenamefont{Ferretti et~al.}(2013)\citenamefont{Ferretti, Galat\`a,
  and Santopinto}}]{Ferretti:2013faa}
\bibinfo{author}{\bibfnamefont{J.}~\bibnamefont{Ferretti}},
  \bibinfo{author}{\bibfnamefont{G.}~\bibnamefont{Galat\`a}}, \bibnamefont{and}
  \bibinfo{author}{\bibfnamefont{E.}~\bibnamefont{Santopinto}},
  \bibinfo{journal}{Phys. Rev. C} \textbf{\bibinfo{volume}{88}},
  \bibinfo{pages}{015207} (\bibinfo{year}{2013}), \eprint{1302.6857}.

\bibitem[{\citenamefont{Liu and Ding}(2012)}]{Liu:2011yp}
\bibinfo{author}{\bibfnamefont{J.-F.} \bibnamefont{Liu}} \bibnamefont{and}
  \bibinfo{author}{\bibfnamefont{G.-J.} \bibnamefont{Ding}},
  \bibinfo{journal}{Eur. Phys. J. C} \textbf{\bibinfo{volume}{72}},
  \bibinfo{pages}{1981} (\bibinfo{year}{2012}), \eprint{1105.0855}.

\bibitem[{\citenamefont{Ferretti et~al.}(2012)\citenamefont{Ferretti, Galata,
  Santopinto, and Vassallo}}]{Ferretti:2012zz}
\bibinfo{author}{\bibfnamefont{J.}~\bibnamefont{Ferretti}},
  \bibinfo{author}{\bibfnamefont{G.}~\bibnamefont{Galata}},
  \bibinfo{author}{\bibfnamefont{E.}~\bibnamefont{Santopinto}},
  \bibnamefont{and} \bibinfo{author}{\bibfnamefont{A.}~\bibnamefont{Vassallo}},
  \bibinfo{journal}{Phys. Rev. C} \textbf{\bibinfo{volume}{86}},
  \bibinfo{pages}{015204} (\bibinfo{year}{2012}).

\bibitem[{\citenamefont{Ferretti and Santopinto}(2014)}]{Ferretti:2013vua}
\bibinfo{author}{\bibfnamefont{J.}~\bibnamefont{Ferretti}} \bibnamefont{and}
  \bibinfo{author}{\bibfnamefont{E.}~\bibnamefont{Santopinto}},
  \bibinfo{journal}{Phys. Rev. D} \textbf{\bibinfo{volume}{90}},
  \bibinfo{pages}{094022} (\bibinfo{year}{2014}), \eprint{1306.2874}.

\bibitem[{\citenamefont{Lu et~al.}(2016)\citenamefont{Lu, Anwar, and
  Zou}}]{Lu:2016mbb}
\bibinfo{author}{\bibfnamefont{Y.}~\bibnamefont{Lu}},
  \bibinfo{author}{\bibfnamefont{M.~N.} \bibnamefont{Anwar}}, \bibnamefont{and}
  \bibinfo{author}{\bibfnamefont{B.-S.} \bibnamefont{Zou}},
  \bibinfo{journal}{Phys. Rev. D} \textbf{\bibinfo{volume}{94}},
  \bibinfo{pages}{034021} (\bibinfo{year}{2016}), \eprint{1606.06927}.

\bibitem[{\citenamefont{van Beveren and Rupp}(2004)}]{vanBeveren:2003jv}
\bibinfo{author}{\bibfnamefont{E.}~\bibnamefont{van Beveren}} \bibnamefont{and}
  \bibinfo{author}{\bibfnamefont{G.}~\bibnamefont{Rupp}},
  \bibinfo{journal}{Eur. Phys. J. C} \textbf{\bibinfo{volume}{32}},
  \bibinfo{pages}{493} (\bibinfo{year}{2004}), \eprint{hep-ph/0306051}.

\bibitem[{\citenamefont{van Beveren and Rupp}(2003)}]{vanBeveren:2003kd}
\bibinfo{author}{\bibfnamefont{E.}~\bibnamefont{van Beveren}} \bibnamefont{and}
  \bibinfo{author}{\bibfnamefont{G.}~\bibnamefont{Rupp}},
  \bibinfo{journal}{Phys. Rev. Lett.} \textbf{\bibinfo{volume}{91}},
  \bibinfo{pages}{012003} (\bibinfo{year}{2003}), \eprint{hep-ph/0305035}.

\bibitem[{\citenamefont{Coito et~al.}(2011)\citenamefont{Coito, Rupp, and van
  Beveren}}]{Coito:2011qn}
\bibinfo{author}{\bibfnamefont{S.}~\bibnamefont{Coito}},
  \bibinfo{author}{\bibfnamefont{G.}~\bibnamefont{Rupp}}, \bibnamefont{and}
  \bibinfo{author}{\bibfnamefont{E.}~\bibnamefont{van Beveren}},
  \bibinfo{journal}{Phys. Rev. D} \textbf{\bibinfo{volume}{84}},
  \bibinfo{pages}{094020} (\bibinfo{year}{2011}), \eprint{1106.2760}.

\bibitem[{\citenamefont{Hwang and Kim}(2004)}]{Hwang:2004cd}
\bibinfo{author}{\bibfnamefont{D.~S.} \bibnamefont{Hwang}} \bibnamefont{and}
  \bibinfo{author}{\bibfnamefont{D.-W.} \bibnamefont{Kim}},
  \bibinfo{journal}{Phys. Lett. B} \textbf{\bibinfo{volume}{601}},
  \bibinfo{pages}{137} (\bibinfo{year}{2004}), \eprint{hep-ph/0408154}.

\bibitem[{\citenamefont{Simonov and Tjon}(2004)}]{Simonov:2004ar}
\bibinfo{author}{\bibfnamefont{Y.~A.} \bibnamefont{Simonov}} \bibnamefont{and}
  \bibinfo{author}{\bibfnamefont{J.~A.} \bibnamefont{Tjon}},
  \bibinfo{journal}{Phys. Rev. D} \textbf{\bibinfo{volume}{70}},
  \bibinfo{pages}{114013} (\bibinfo{year}{2004}), \eprint{hep-ph/0409361}.

\bibitem[{\citenamefont{Lee et~al.}(2007)\citenamefont{Lee, Lee, Min, and
  Park}}]{Lee:2004gt}
\bibinfo{author}{\bibfnamefont{I.~W.} \bibnamefont{Lee}},
  \bibinfo{author}{\bibfnamefont{T.}~\bibnamefont{Lee}},
  \bibinfo{author}{\bibfnamefont{D.~P.} \bibnamefont{Min}}, \bibnamefont{and}
  \bibinfo{author}{\bibfnamefont{B.-Y.} \bibnamefont{Park}},
  \bibinfo{journal}{Eur. Phys. J. C} \textbf{\bibinfo{volume}{49}},
  \bibinfo{pages}{737} (\bibinfo{year}{2007}), \eprint{hep-ph/0412210}.

\bibitem[{\citenamefont{Guo et~al.}(2008)\citenamefont{Guo, Krewald, and
  Meissner}}]{Guo:2007up}
\bibinfo{author}{\bibfnamefont{F.-K.} \bibnamefont{Guo}},
  \bibinfo{author}{\bibfnamefont{S.}~\bibnamefont{Krewald}}, \bibnamefont{and}
  \bibinfo{author}{\bibfnamefont{U.-G.} \bibnamefont{Meissner}},
  \bibinfo{journal}{Phys. Lett. B} \textbf{\bibinfo{volume}{665}},
  \bibinfo{pages}{157} (\bibinfo{year}{2008}), \eprint{0712.2953}.

\bibitem[{\citenamefont{Zhou and Xiao}(2011)}]{Zhou:2011sp}
\bibinfo{author}{\bibfnamefont{Z.-Y.} \bibnamefont{Zhou}} \bibnamefont{and}
  \bibinfo{author}{\bibfnamefont{Z.}~\bibnamefont{Xiao}},
  \bibinfo{journal}{Phys. Rev. D} \textbf{\bibinfo{volume}{84}},
  \bibinfo{pages}{034023} (\bibinfo{year}{2011}), \eprint{1105.6025}.

\bibitem[{\citenamefont{Badalian et~al.}(2008)\citenamefont{Badalian, Simonov,
  and Trusov}}]{Badalian:2007yr}
\bibinfo{author}{\bibfnamefont{A.~M.} \bibnamefont{Badalian}},
  \bibinfo{author}{\bibfnamefont{Y.~A.} \bibnamefont{Simonov}},
  \bibnamefont{and} \bibinfo{author}{\bibfnamefont{M.~A.}
  \bibnamefont{Trusov}}, \bibinfo{journal}{Phys. Rev. D}
  \textbf{\bibinfo{volume}{77}}, \bibinfo{pages}{074017}
  (\bibinfo{year}{2008}), \eprint{0712.3943}.

\bibitem[{\citenamefont{Dai et~al.}(2008)\citenamefont{Dai, Li, Zhu, and
  Zuo}}]{Dai:2006uz}
\bibinfo{author}{\bibfnamefont{Y.-B.} \bibnamefont{Dai}},
  \bibinfo{author}{\bibfnamefont{X.-Q.} \bibnamefont{Li}},
  \bibinfo{author}{\bibfnamefont{S.-L.} \bibnamefont{Zhu}}, \bibnamefont{and}
  \bibinfo{author}{\bibfnamefont{Y.-B.} \bibnamefont{Zuo}},
  \bibinfo{journal}{Eur. Phys. J. C} \textbf{\bibinfo{volume}{55}},
  \bibinfo{pages}{249} (\bibinfo{year}{2008}), \eprint{hep-ph/0610327}.

\bibitem[{\citenamefont{Ferretti and Santopinto}(2018)}]{Ferretti:2015rsa}
\bibinfo{author}{\bibfnamefont{J.}~\bibnamefont{Ferretti}} \bibnamefont{and}
  \bibinfo{author}{\bibfnamefont{E.}~\bibnamefont{Santopinto}},
  \bibinfo{journal}{Phys. Rev. D} \textbf{\bibinfo{volume}{97}},
  \bibinfo{pages}{114020} (\bibinfo{year}{2018}), \eprint{1506.04415}.

\bibitem[{\citenamefont{Silvestre-Brac and
  Gignoux}(1991)}]{Silvestre-Brac:1991qqx}
\bibinfo{author}{\bibfnamefont{B.}~\bibnamefont{Silvestre-Brac}}
  \bibnamefont{and} \bibinfo{author}{\bibfnamefont{C.}~\bibnamefont{Gignoux}},
  \bibinfo{journal}{Phys. Rev. D} \textbf{\bibinfo{volume}{43}},
  \bibinfo{pages}{3699} (\bibinfo{year}{1991}).

\bibitem[{\citenamefont{Geiger and Isgur}(1991{\natexlab{a}})}]{Geiger:1991ab}
\bibinfo{author}{\bibfnamefont{P.}~\bibnamefont{Geiger}} \bibnamefont{and}
  \bibinfo{author}{\bibfnamefont{N.}~\bibnamefont{Isgur}},
  \bibinfo{journal}{Phys. Rev. D} \textbf{\bibinfo{volume}{44}},
  \bibinfo{pages}{799} (\bibinfo{year}{1991}{\natexlab{a}}).

\bibitem[{\citenamefont{Geiger and Isgur}(1991{\natexlab{b}})}]{Geiger:1991qe}
\bibinfo{author}{\bibfnamefont{P.}~\bibnamefont{Geiger}} \bibnamefont{and}
  \bibinfo{author}{\bibfnamefont{N.}~\bibnamefont{Isgur}},
  \bibinfo{journal}{Phys. Rev. Lett.} \textbf{\bibinfo{volume}{67}},
  \bibinfo{pages}{1066} (\bibinfo{year}{1991}{\natexlab{b}}).

\bibitem[{\citenamefont{Geiger and Isgur}(1997)}]{Geiger:1996re}
\bibinfo{author}{\bibfnamefont{P.}~\bibnamefont{Geiger}} \bibnamefont{and}
  \bibinfo{author}{\bibfnamefont{N.}~\bibnamefont{Isgur}},
  \bibinfo{journal}{Phys. Rev. D} \textbf{\bibinfo{volume}{55}},
  \bibinfo{pages}{299} (\bibinfo{year}{1997}), \eprint{hep-ph/9610445}.

\bibitem[{\citenamefont{Ackleh et~al.}(1996)\citenamefont{Ackleh, Barnes, and
  Swanson}}]{Ackleh:1996yt}
\bibinfo{author}{\bibfnamefont{E.~S.} \bibnamefont{Ackleh}},
  \bibinfo{author}{\bibfnamefont{T.}~\bibnamefont{Barnes}}, \bibnamefont{and}
  \bibinfo{author}{\bibfnamefont{E.~S.} \bibnamefont{Swanson}},
  \bibinfo{journal}{Phys. Rev. D} \textbf{\bibinfo{volume}{54}},
  \bibinfo{pages}{6811} (\bibinfo{year}{1996}), \eprint{hep-ph/9604355}.

\bibitem[{\citenamefont{Barnes et~al.}(2005)\citenamefont{Barnes, Godfrey, and
  Swanson}}]{Barnes:2005pb}
\bibinfo{author}{\bibfnamefont{T.}~\bibnamefont{Barnes}},
  \bibinfo{author}{\bibfnamefont{S.}~\bibnamefont{Godfrey}}, \bibnamefont{and}
  \bibinfo{author}{\bibfnamefont{E.~S.} \bibnamefont{Swanson}},
  \bibinfo{journal}{Phys. Rev. D} \textbf{\bibinfo{volume}{72}},
  \bibinfo{pages}{054026} (\bibinfo{year}{2005}), \eprint{hep-ph/0505002}.

\bibitem[{\citenamefont{Chen et~al.}(2018)\citenamefont{Chen, Ping, Roberts,
  and Segovia}}]{Chen:2017mug}
\bibinfo{author}{\bibfnamefont{X.}~\bibnamefont{Chen}},
  \bibinfo{author}{\bibfnamefont{J.}~\bibnamefont{Ping}},
  \bibinfo{author}{\bibfnamefont{C.~D.} \bibnamefont{Roberts}},
  \bibnamefont{and} \bibinfo{author}{\bibfnamefont{J.}~\bibnamefont{Segovia}},
  \bibinfo{journal}{Phys. Rev. D} \textbf{\bibinfo{volume}{97}},
  \bibinfo{pages}{094016} (\bibinfo{year}{2018}), \eprint{1712.04457}.

\bibitem[{\citenamefont{Ortega et~al.}(2016)\citenamefont{Ortega, Segovia,
  Entem, and Fernandez}}]{Ortega:2016mms}
\bibinfo{author}{\bibfnamefont{P.~G.} \bibnamefont{Ortega}},
  \bibinfo{author}{\bibfnamefont{J.}~\bibnamefont{Segovia}},
  \bibinfo{author}{\bibfnamefont{D.~R.} \bibnamefont{Entem}}, \bibnamefont{and}
  \bibinfo{author}{\bibfnamefont{F.}~\bibnamefont{Fernandez}},
  \bibinfo{journal}{Phys. Rev. D} \textbf{\bibinfo{volume}{94}},
  \bibinfo{pages}{074037} (\bibinfo{year}{2016}), \eprint{1603.07000}.

\bibitem[{\citenamefont{Mart\'\i{}nez~Torres
  et~al.}(2015)\citenamefont{Mart\'\i{}nez~Torres, Oset, Prelovsek, and
  Ramos}}]{MartinezTorres:2014kpc}
\bibinfo{author}{\bibfnamefont{A.}~\bibnamefont{Mart\'\i{}nez~Torres}},
  \bibinfo{author}{\bibfnamefont{E.}~\bibnamefont{Oset}},
  \bibinfo{author}{\bibfnamefont{S.}~\bibnamefont{Prelovsek}},
  \bibnamefont{and} \bibinfo{author}{\bibfnamefont{A.}~\bibnamefont{Ramos}},
  \bibinfo{journal}{JHEP} \textbf{\bibinfo{volume}{05}}, \bibinfo{pages}{153}
  (\bibinfo{year}{2015}), \eprint{1412.1706}.

\bibitem[{\citenamefont{Song et~al.}(2023)\citenamefont{Song, Dai, and
  Oset}}]{Song:2023pdq}
\bibinfo{author}{\bibfnamefont{J.}~\bibnamefont{Song}},
  \bibinfo{author}{\bibfnamefont{L.~R.} \bibnamefont{Dai}}, \bibnamefont{and}
  \bibinfo{author}{\bibfnamefont{E.}~\bibnamefont{Oset}}
  (\bibinfo{year}{2023}), \eprint{2307.02382}.

\bibitem[{\citenamefont{Dai et~al.}(2023)\citenamefont{Dai, Song, and
  Oset}}]{Dai:2023kwv}
\bibinfo{author}{\bibfnamefont{L.~R.} \bibnamefont{Dai}},
  \bibinfo{author}{\bibfnamefont{J.}~\bibnamefont{Song}}, \bibnamefont{and}
  \bibinfo{author}{\bibfnamefont{E.}~\bibnamefont{Oset}},
  \bibinfo{journal}{Phys. Lett. B} \textbf{\bibinfo{volume}{846}},
  \bibinfo{pages}{138200} (\bibinfo{year}{2023}), \eprint{2306.01607}.

\bibitem[{\citenamefont{Godfrey and Moats}(2014)}]{Godfrey:2014fga}
\bibinfo{author}{\bibfnamefont{S.}~\bibnamefont{Godfrey}} \bibnamefont{and}
  \bibinfo{author}{\bibfnamefont{K.}~\bibnamefont{Moats}},
  \bibinfo{journal}{Phys. Rev. D} \textbf{\bibinfo{volume}{90}},
  \bibinfo{pages}{117501} (\bibinfo{year}{2014}), \eprint{1409.0874}.

\bibitem[{\citenamefont{Aubert et~al.}(2009)}]{BaBar:2009rro}
\bibinfo{author}{\bibfnamefont{B.}~\bibnamefont{Aubert}} \bibnamefont{et~al.}
  (\bibinfo{collaboration}{BaBar}), \bibinfo{journal}{Phys. Rev. D}
  \textbf{\bibinfo{volume}{80}}, \bibinfo{pages}{092003}
  (\bibinfo{year}{2009}), \eprint{0908.0806}.

\end{thebibliography}

\end{document}